\documentstyle[12pt,cite]{article}

\newcommand{\be}{\begin{equation}}
\newcommand{\ee}{\end{equation}}
\def\bea{\begin{eqnarray}}
\def\eea{\end{eqnarray}}
\def\nn{\nonumber \\}
\def\qqg{$q {\bar q}g$}
\def\qq{$q {\bar q}$}
\def\pp{$p_{\perp}$}
\def\x{x_{\perp}}
\def\y{y_{\perp}}

\def\l{l'_{\perp}}

\def\0{0_{\perp}}
\def\lapprox{\lower .7ex\hbox{$\;\stackrel{\textstyle <}{\sim}\;$}}
\def\gapprox{\lower .7ex\hbox{$\;\stackrel{\textstyle >}{\sim}\;$}}

\input psfig

\begin{document}
\date{}
\title{
{\large\rm DESY 99-076}\hfill{\large\tt ISSN 0418-9833}\\
{\large\rm June 1999}
\hfill\vspace*{2.5cm}\\
{\bf Towards the Theory of Diffractive DIS}\footnote{Presented at {\it New 
Trends in HERA Physics 1999}, Ringberg Workshop, June 1999}}
\author{W. Buchm\"uller\\
{\normalsize\it Deutsches Elektronen-Synchrotron DESY, 22603 Hamburg, 
Germany}
\vspace*{2cm}\\                     
}

\maketitle  
\begin{abstract}
\noindent
The large rapidity gap events, observed at HERA, have changed considerably
our physical picture of deep inelastic scattering during the past years. 
We review the present theoretical understanding of diffractive DIS with
emphasis on the close relation to inclusive DIS. This includes success and
limitations of the leading twist description, the connection between
diffractive and inclusive parton distributions in the semiclassical
approach, the colour structure of the proton and comparison with data.
The progress report concludes with a list of open questions.
\end{abstract}
\thispagestyle{empty}
\newpage                                             

\section{Inclusive and diffractive DIS}

The intriguing phenomenon of the frequent appearance of large rapidity gaps
in electron proton collisions at HERA \cite{h1z} has changed our physical 
picture of deep inelastic scattering (DIS) to a large extent. The large 
rapidity gap events are very difficult to understand in the parton model 
where the struck quark is expected to break up the proton leading to a 
continuous flow of hadrons between the current jet and the proton remnant.

To develop a physical picture of diffractive DIS \cite{heb} it is convenient 
to view the scattering process in the proton rest frame. In this frame the
virtual photon fluctuates into partonic states \qq , \qqg , ... which then
scatter off the proton. From the leading twist contributions to inclusive and
diffractive structure functions one obtains the parton distribution functions
in a frame where the proton moves fast. This connection holds for diffractive
as well as non-diffractive processes.

In the following we shall review the present status of our theoretical
understanding of diffractive DIS with emphasis on the close analogy to
inclusive DIS. This is appropriate since diffractive DIS is dominated by
the leading twist contribution, which has been one of the most surprising 
aspects of the large rapidity gap events. The scattering of the partonic
fluctuations of the photon off the proton will be treated in the
semiclassical approach. After a comparison of theoretical predictions
with data we shall conclude with a discussion of some open questions.   
 
\subsection*{Inclusive DIS}

Inclusive deep inelastic scattering \cite{esw} is characterized by the 
kinematic variables
\be
Q^2=-q^2\;,\quad W^2=(q+P)^2\;, \quad x={Q^2\over Q^2 + W^2}\;,
\ee
where $q$ and $P$ are the momenta of the virtual photon and the proton,
respectively. The cross section is determined by the hadronic tensor,
\bea
W_{\mu\nu}(P,q) &=& {1\over 4\pi} \sum_{X} 
  \langle P|J_\nu(0)|X\rangle  
  \langle X|J_\mu(0)|P\rangle (2\pi)^4 \delta(P-P_X)\nonumber\\
&=&\left(-g_{\mu\nu}+{q_\mu q_\nu\over q^2}\right) F_1(x,Q^2)
  +{1\over \nu}\left(P_\mu - {\nu\over q^2}q_\mu\right) 
   \left(P_\nu - {\nu\over q^2}q_\nu\right) F_2(x,Q^2)\;.
\eea
Here $J_\mu(x)$ is the electromagnetic current, $\nu=q\cdot P$, and 
spin averaging has been implicitly assumed.

The structure functions are a sum of leading twist and of higher twist
contributions which are suppressed by powers of $Q^2$,
\be
F_i(x,Q^2) = F_i^{(LT)}(x,Q^2) + {F_i^{(HT)}(x,Q^2)\over Q^2} + \ldots\;.
\ee
The leading twist term is dominant for $Q^2$ above some value  
$Q_0^2$, which is not very well known and frequently chosen to be 
${\cal O}$(1 GeV$^2$). However, higher twist contributions are known to
be important for hadronic energies $W^2 \le 4$~GeV$^2$ \cite{mrst2}.

The structure functions $F_i^{(LT)}(x,Q^2)$ can be expressed in terms of
process independent parton distribution functions,
\be
F_i^{(LT)}(x,Q^2) \rightarrow f_i(x,\mu^2) = q(x,\mu^2),\ g(x,\mu^2)\;,
\ee
which depend on $x$ and on the factorization scale $\mu^2$. At small $x$, the
quark distribution is assumed to be the same for all light flavours.
The parton distribution functions $f_i(x,\mu^2)$ obey the perturbative
QCD evolution equations \cite{dglap},
\be
\mu^2 {\partial\over \partial\mu^2} f_i(x,\mu^2) =
{\alpha_s\over 2\pi} \int_x^1{dy\over y} P_{ij}\left({x\over y}\right)
f_i(y,\mu^2)\;,
\ee
where $P_{ij}(z)$ are the Altarelli-Parisi splitting functions. 
The parton distributions can be directly expressed in terms of
the quark and gluon field operators. For instance, the quark distribution
is given by
\bea\label{qdincl}
q(x,\mu^2) &=& {1\over 4\pi} \int dx_- e^{-ixP_+x_-/2}\sum_{X}\nonumber\\
&&\hspace{1.2cm}
\langle P|\bar{q}(0,x_-,\0)U(x_-,\infty)|X\rangle\gamma_+
\langle X|U(\infty,0)q(0,0,\0)|P\rangle \;,
\eea
where $U(a,b)$ is the colour matrix
\be
U(a,b) = P \exp{\left(-{i\over 2}\int_b^a dy_-A_+(0,y_-,\0)\right)}\;.
\ee
This definition can be used as a starting point of a theoretical
non-perturbative evaluation of the quark distribution.

\subsection*{Diffractive DIS}

Diffractive DIS can be discussed in close analogy to inclusive DIS. There 
are two more kinematical variables which charaterize the diffractively 
scattered proton: the invariant momentum transfer $t$ and the fraction 
$\xi$ of lost longitudinal momentum. A complete set
of variables is
\be
t=(P-P')^2\;,\quad \xi\equiv x_{I\!\!P} \;,\quad Q^2=-q^2\;,\quad 
M^2=(q+\xi P)^2\;,\quad \beta={Q^2\over Q^2 + M^2}\;.
\ee
Compared to inclusive DIS, the diffractive mass $M$ plays the role of the
total hadronic mass $W$, and $\beta$ corresponds to $x$.

The hadronic tensor for diffractive DIS,
\bea 
W^D_{\mu\nu}(P,P',q) &=& {1\over 4\pi} \sum_{X} 
  \langle P|J_\nu(0)|X;P'\rangle  
  \langle X;P'|J_\mu(0)|P\rangle 
  (2\pi)^4 \delta(P-P'-P_X)\nonumber\\
&=&\left(-g_{\mu\nu}+{q_\mu q_\nu\over q^2}\right) F_1^{D(4)}(t,\xi,\beta,Q^2)
   \nonumber\\ 
&&\quad+{1\over \nu}\left(P_\mu - {\nu\over q^2}q_\mu\right) 
\left(P_\nu - {\nu\over q^2}q_\nu\right) F_2^{D(4)}(t,\xi,\beta,Q^2)+\ldots\;,
\eea
defines the diffractive structure functions $F_i^{D(4)}(t,\xi,\beta,Q^2)$.
Integration over $t$, which is dominated by small $|t|$ for diffractive
scattering, yields the extensively studied structure function
\be
F_2^{D(3)}(\xi,\beta,Q^2) = \int dt F_2^{D(4)}(t,\xi,\beta,Q^2)\;.
\ee

Also the diffractive structure functions have contributions of leading and
higher twist, 
\be
F_i^{D(3)}(\xi,\beta,Q^2) = F_i^{D(3,LT)}(\xi,\beta,Q^2) + 
{F_i^{D(3,HT)}(\xi,\beta,Q^2)\over Q^2} + \ldots\;.
\ee
Again it is unclear above which value of $Q^2_0$ the leading twist part
dominates. At small $x$, $W^2\simeq Q^2/x$ should be large enough, whereas
the lower bound on $M^2$ is an open question. Our phenomenological analysis
in the next section will show that the leading twist description breaks
down at $M_0^2 \simeq 4$ GeV$^2$. This again demonstrates that in diffractive
DIS $M^2$ plays a role analogous to $W^2$ in inclusive DIS.  

For diffractive DIS factorization holds like for inclusive DIS \cite{c}. 
The diffractive structure functions $F_i^{D(3,LT)}(\xi,\beta,Q^2)$ can be 
expressed in terms of `fracture functions' \cite{tre}, or
`diffractive parton distributions' \cite{ber1},
\be
F_i^{D(3,LT)}(\xi,\beta,Q^2) \rightarrow {df_i(\xi,\beta,\mu^2)\over d\xi} = 
{dq(\xi,\beta,\mu^2)\over d\xi}\;,\; {dg(\xi,\beta,\mu^2)\over d\xi}\;,
\ee
which depend on $\xi$, $\beta$ and the factorization scale $\mu^2$. 
The diffractive parton distribution functions $df_i(\xi,\beta,\mu^2)/d\xi$ 
also obey the perturbative QCD evolution equations,
\be\label{ddglap}
\mu^2 {\partial\over \partial\mu^2} {df_i(\xi,\beta,\mu^2)\over d\xi} =
{\alpha_s\over 2\pi} \int_\beta^1{db\over b} P_{ij}\left({\beta\over b}\right)
{df_i(\xi,b,\mu^2)\over d\xi}\;.
\ee
Note that the evolution takes now place in $\beta$ and $Q^2$; $\xi$ 
merely acts as a parameter. The physical reason for this is intuitively clear: 
for an arbitrary DIS event the invariant hadronic mass is $W$, and the quark 
which couples to the virtual photon can be radiated by a parton whose fraction
of the proton momentum varies from 1 to $x=Q^2/(Q^2+W^2)$. In a diffractive 
event, the diffractive invariant mass is $M$. Hence, $W$ is 
replaced by $M$, and the quark which couples to the photon can be radiated 
by a parton whose fraction of the momentum $\xi P$ varies from 1 to 
$\beta=Q^2/(Q^2+M^2)$. Formally, Eq.~(\ref{ddglap}) follows from the fact 
that ultraviolet divergencies and renormalization are the same for  
inclusive and diffractive parton distribution functions \cite{ber2}. This is 
apparent from a comparison of the corresponding operator definitions. The 
diffractive quark distribution, for instance, is given by \cite{ber2},
\bea\label{qddiff}
{dq(\xi,\beta,\mu^2)\over d\xi} 
&=& {1\over 64\pi^3}\int dt \int dx_- e^{-ixP_+x_-/2}\sum_{X}\nonumber\\
&&\langle P|\bar{q}(0,x_-,\vec{0})U(x_-,\infty)|X;P'\rangle\gamma_+
\langle X;P'|U(\infty,0)q(0,0,\vec{0})|P\rangle \;.
\eea
Assuming `Regge factorization' for the diffractive quark and gluon distribution
functions yields the Ingelman-Schlein model of hard diffractive scattering
\cite{ing} which can also be applied to deep inelastic scattering \cite{dl}.

The physical interpretation of the diffractive parton distributions is 
analogous to the interpretation of the inclusive distributions. The function 
$df(\xi,b,\mu^2)/d\xi$ is a conditional probability distribution. It describes
the probability density to find a parton $f$, carrying a fraction $\xi b$ of 
the proton momentum, under the condition that the proton has lost a fraction 
$\xi$ of its momentum in the scattering process. 

The formal definition of diffractive parton distributions tells us very
little about their properties, although, comparing Eqs.~(\ref{qdincl}) and
(\ref{qddiff}), one may expect that diffractive DIS is a leading twist
effect. However, the important physics question concerns the relation
between the two types of distribution functions,
\be
f_i(x,\mu^2) \longleftrightarrow {df_i(\xi,\beta,\mu^2)\over d\xi}\quad ?
\ee 
Both kinds of parton distributions represent non-perturbative properties
of the proton and are therefore not accessible to perturbation theory.
Still, one may hope that at small $x$, i.e. large hadronic energies $W$,
some simple relations between inclusive and diffractive deep inelastic
scattering may exist. In the following section we shall describe a picture
of hadrons at small $x$ where this is indeed the case.

\section{Semiclassical approach}

The phenomenon of the large rapidity gap events in DIS is very difficult
to understand within the parton model. Naively, one would expect that the
struck quark will always break up the proton, which should lead to a flow
of hadrons between the current jet and the proton remnant without large
gaps in rapidity. 

\begin{figure}
\begin{center}
\vspace*{-.5cm}
\parbox[b]{13.7cm}{\psfig{width=10cm,file=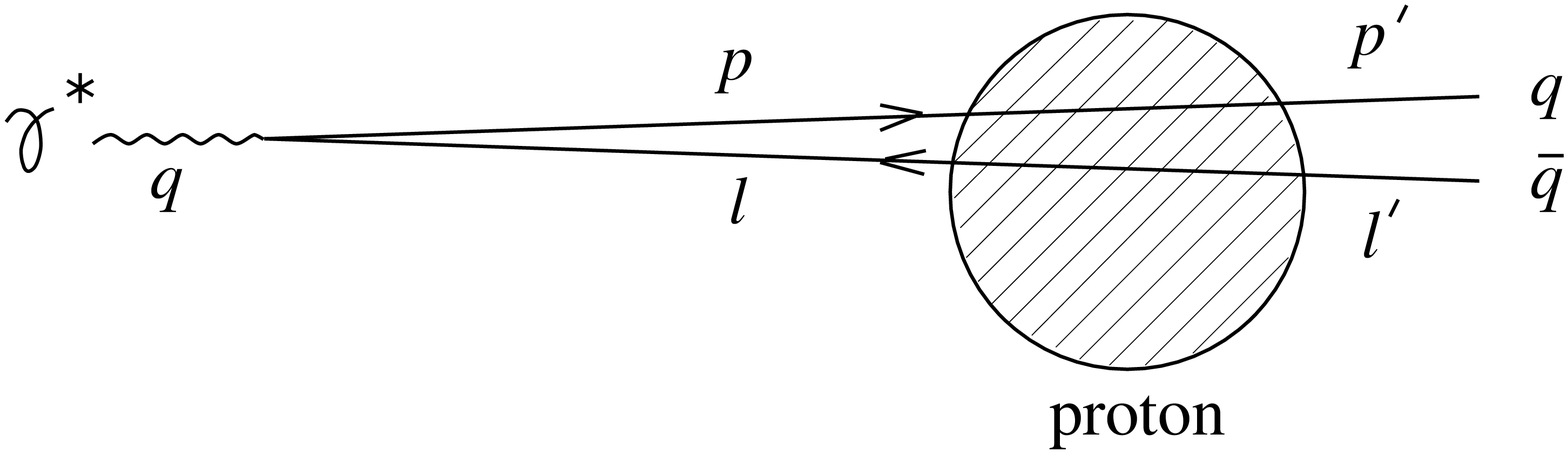}}\\
\end{center}
\refstepcounter{figure}
\label{f1}
{\bf Figure \ref{f1}:} Diffractive or non-diffractive DIS in the proton rest 
frame; the proton is viewed as a superposition of colour fields with
size $1/\Lambda$. 
\end{figure}

The connection between diffractive DIS and ordinary, non-diffractive DIS can
be most easily understood in the proton rest frame which has frequently been
used in the early days of DIS, almost 30 years ago. In this frame, DIS 
appears as the scattering of partonic fluctuations of the photon, \qq, \qqg\ 
etc., off the proton. In the semiclassical approach \cite{bh1} the proton is 
viewed as a superposition of colour fields of size $1/\Lambda$ in DIS at small
$x$, i.e. at high $\gamma^* p$ center-of-mass energies. The simplest partonic
fluctuation is a quark-antiquark pair (cf.~Fig.~1). Penetrating the proton,
quark and antiquark change their colour. If the \qq\ pair leaves the proton 
in a colour singlet configuration, it can fragment independently of the
proton remnant yielding a diffractive event. A \qq\ pair in a colour octet
state will build up a flux tube together with the proton remnant, whose
breakup will lead to an ordinary non-diffractive event.  

The scattering amplitude for both types of events is determined by a single
non-perturbative quantity, $\mbox{tr}W_{\x}(\y)$. Here $\x$ and 
$\x+\y$ are the transverse positions where quark and antiquark penetrate the 
colour field of the proton. The function 
\be
W_{\x}(\y)=U(\x)U^{\dagger}(\x+\y)-1\;, \label{wuu}
\ee
with
\be
U(\x) = P \exp{\left(-{i\over 2}\int_{-\infty}^{\infty} 
        dx_-A_+(0,x_-,\x)\right)}\;,
\ee
is essentially a closed Wilson loop through the corresponding section of 
the proton, which measures an integral of the proton colour field strength. 

Diffractive DIS requires a colour singlet pair in the final state. Hence the
scattering amplitude is $\propto \mbox{tr}W_{\x}(\y)$ and the diffractive
cross section takes the form,
\be\label{dsdiff}
d\sigma^D \propto \int_{\x} \ldots |\ldots \mbox{tr}W_{\x}(\y)\ldots|^2\;.
\ee 
The inclusive cross section is obtained by summing over all colours, which
yields
\bea
d\sigma^{incl} &\propto& \int_{\x} \ldots \mbox{tr}
        \left(W_{\x}(\y)W^\dagger_{\x}(\y)\right)\ldots \nonumber\\
&\propto& \int_{\x} \ldots \mbox{tr}W_{\x}(\y)\ldots \;,\label{dsincl}
\eea 
where the last equation follows from the unitarity of the matrix $U(\x)$.

From Eqs.~(\ref{dsdiff}) and (\ref{dsincl}) one immediately derives the
properties of Bjorken's aligned jet model \cite{bjo}. For small 
quark-antiquark separations one has,
\be
\int_{\x} \mbox{tr}W_{\x}(\y) \propto \y^2 \;.
\ee
Hence, since all kinematical factors are the same for $d\sigma^D$ and
$d\sigma^{incl}$, small \qq\ pairs are suppressed in diffractive DIS. For
large pairs of size $1/\Lambda$, the transverse momentum $\l$ and the
longitudinal momentum fractions $\alpha$ and $1-\alpha$ are
\be
\l \sim \Lambda\;, \quad \alpha \sim {\Lambda^2\over Q^2}\;, 
\quad 1-\alpha \simeq 1\;.
\ee
These are the asymmetric, aligned jet configurations \cite{bjo} which
dominate diffractive DIS. 

\begin{figure}
\begin{center}
\vspace*{-.5cm}
\parbox[b]{15cm}{\psfig{width=15cm,file=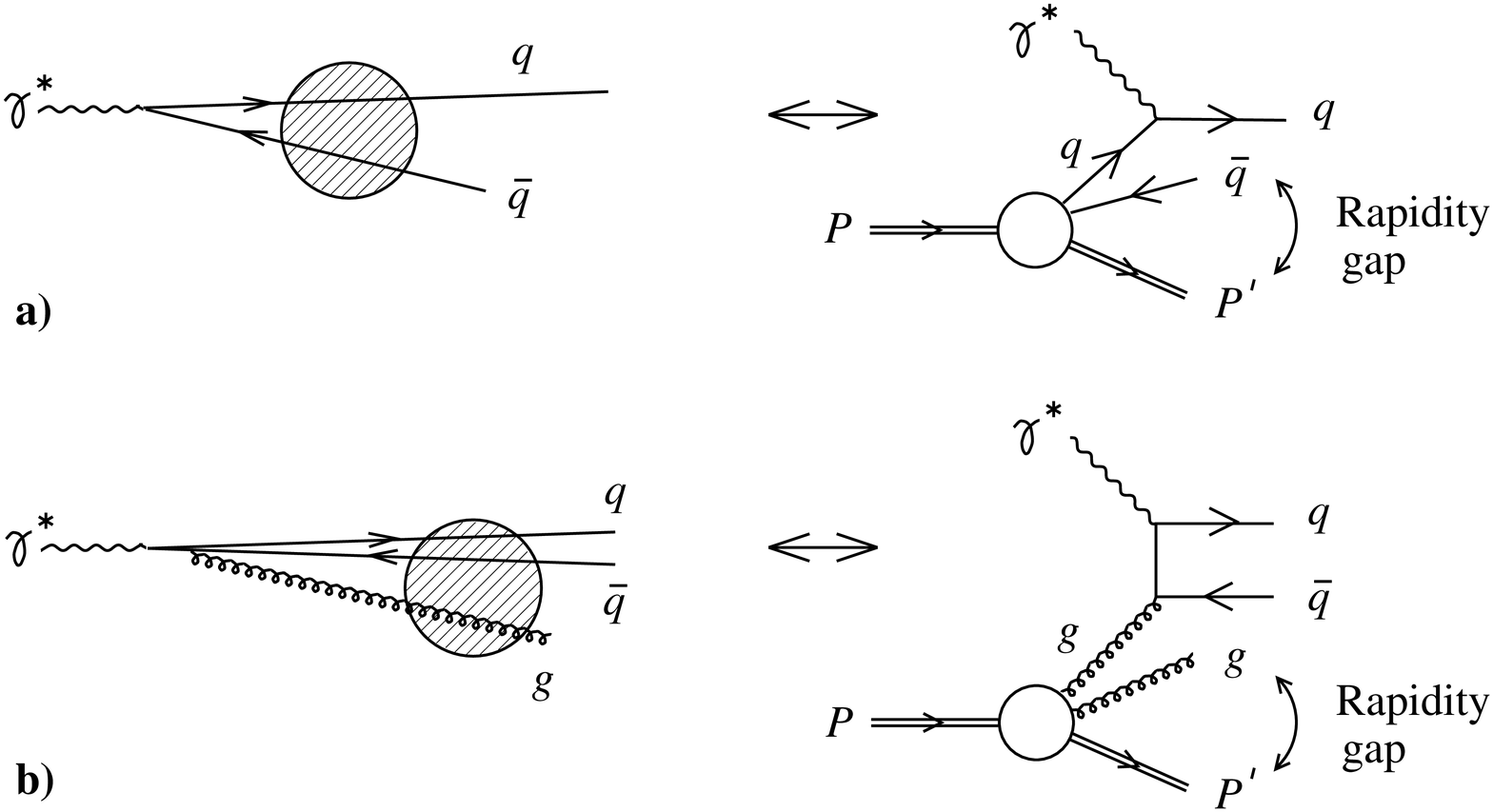}}\\
\end{center}
\refstepcounter{figure}
\label{f2d}
{\bf Figure \ref{f2d}:} Diffractive DIS in the proton rest frame (left) and 
the Breit frame (right); asymmetric quark fluctuations correspond to 
diffractive quark scattering, asymmetric gluon fluctuations to diffractive 
boson-gluon fusion. 
\begin{center}
\vspace*{-.5cm}
\parbox[b]{13.7cm}{\psfig{width=13.7cm,file=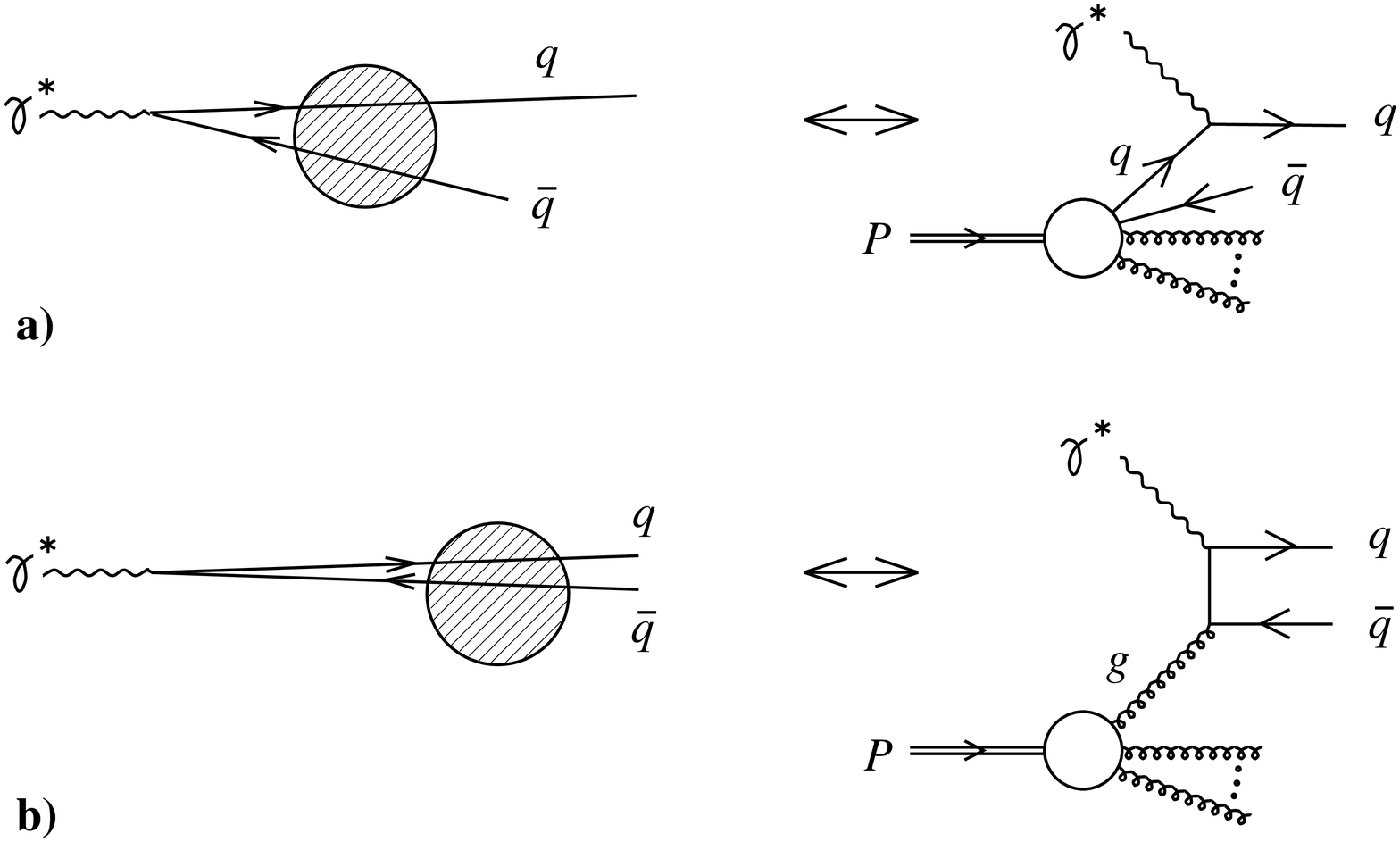}}\\
\end{center}
\refstepcounter{figure}
\label{f2}
{\bf Figure \ref{f2}:} Inclusive DIS in the proton rest frame (left) and the 
Breit frame (right); asymmetric fluctuations correspond to quark scattering 
(a), symmetric fluctuations to boson-gluon fusion (b). 
\end{figure}

\subsection*{Diffractive and inclusive parton distributions}

In the semiclassical approach the evaluation of inclusive and diffractive 
structure functions is straightforward, in principle. One has to calculate
the scattering amplitudes for the production of  \qq\ , \qqg\ ... 
configurations \cite{bhm} in an external colour field, analogous to the 
production of $\mu^+\mu^-$ pairs in an external electromagnetic field 
\cite{bks}, treat the interaction of the fast partons with the non-abelian 
colour field in the eikonal approximation \cite{n}, and finally integrate 
over all target colour fields.

The result for the leading twist part can be expressed in terms of
diffractive parton distributions \cite{h}. For the transverse structure
function, for instance, one finds to leading order in the QCD coupling,
\bea
F_T^D(\xi,\beta,Q^2) &=& 2e_q^2x \int_{\beta}^1 {db\over b} \Bigg\{\left(
\delta(1-z) + {\alpha_s\over 2\pi}\left(P_{qq}(z)\ln{Q^2\over \mu^2} + 
\ldots\right) \right){dq(b,\xi,\mu^2)\over d\xi}\nonumber\\
&&\hspace{1cm} + {\alpha_s\over 2\pi}\left(P_{qg}(z)\ln{Q^2\over \mu^2}
+\ldots\right){dg(b,\xi,\mu^2)\over d\xi}\Bigg\}\;,\label{ftd}
\eea
where $z=\beta/b$, and $C_F$ and $T_F$ are the usual colour factors. This
expression is completely analogous to the well known result for the
inclusive structure function $F_T(x,Q^2)$. In the diffractive case, 
$\beta$ plays the role of $x$, whereas $\xi$ only acts as a parameter.
From  Eq.~(\ref{ftd}) it is obvious that, as anticipated, the diffractive 
parton distributions satisfy the perturbative QCD evolution equations 
(\ref{ddglap}).

The diffractive quark and gluon distributions have been determined in 
\cite{h}. In terms of Wilson loops in coordinate space, the quark 
distribution can be expressed as follows, 
\bea\label{dqd}
{dq(\xi,b,\mu^2)\over d\xi}&=&{2b\over \xi^2(1-b)^3}
           \int{d^2\l {\l}^4\over(2\pi)^6 N_c} 
           \int_{\y,\y'} e^{i\l(\y-\y')}\,{\y\y'\over y\, y'}\nn\nn
&&\times\,K_1(yN)K_1(y'N)\int_{\x}\mbox{tr}W_{\x}(\y)
\mbox{tr}W^{\dagger}_{\x}(\y')\;,  
\eea
where $N_c$ is the number of colours and $N^2={\l}^2{b\over 1-b}$.

It is very instructive to compare diffractive DIS in the proton rest frame
and in the Breit frame (cf.~Fig. 2). The number of partons in the final
state is the same, of course, in both frames. Note, however, that the virtual
parton connected to the proton changes it's direction. It appears incoming
in the proton rest frame and outgoing in the Breit frame. Diffractive quark
and gluon distributions correspond to asymmetric \qq\ and \qqg\ fluctuations
with a slow antiquark and gluon, respectively.

Inclusive parton distributions can be calculated in a similar way. The
inclusive quark distribution is again given by the asymmetric \qq\ 
configuration (cf.~Fig.~3), just with arbitrary colours in the final state.
A special role is played by the inclusive gluon distribution. It
is related to small symmetric \qq\ pairs which probe the colour field of the 
proton directly (cf.~Fig.~3). Contrary to all other parton distributions,
the inclusive gluon distribution \cite{bgh},

\bea
xg(x,Q^2) &=& \frac{3\pi}{\alpha_s e_q^2}\,\cdot\,\frac{\partial F_T(x,Q^2)}
{\partial\ln Q^2} \label{ctrans}\\
&=& {1\over 2\pi^2 \alpha_s}\int_{\x} \mbox{tr}\left(
\partial_{\y}W_{\x}(0)\partial_{\y}W_{\x}^{\dagger}(0)\right)
= {\cal O}\left({1\over \alpha_s}\right)\;,\label{gi}
\eea
is enhanced by an inverse power of $\alpha_s$ in the semiclassical approach.
This is the reason why diffractive DIS is suppressed. Note that the gluon 
distribution is directly related to the cross section for a small
\qq\ pair with transverse size $y$ \cite{fms},
\be
\sigma_{q\bar{q}} (y;x,Q^2) = {\pi^2\over 3} \alpha_s x g(x,Q^2) y^2 + 
{\cal O}(y^4)\;.
\ee

\subsection*{Integration over the target gluon fields}\label{av}

So far we have expressed diffractive and inclusive parton distributions in 
terms of Wilson loops which integrate the gluon field strength in the area
between the trajectories of two fast colour charges penetrating the proton.
The integration over the gluon field configurations of the target is a 
complicated operation depending on the full details of the non-perturbative 
hadronic state. However, in the special case of a very large target, a 
quantitative treatment becomes possible under minimal additional assumptions.
The reason is that the large size of a hadronic target, realized, e.g., in an 
extremely heavy nucleus, introduces a new hard scale \cite{mv}. From the 
target rest frame point of view, this means that the 
typical transverse size of the partonic fluctuations of the virtual photon 
remains perturbative \cite{hw}, thus justifying the omission of higher Fock 
states in the semiclassical calculation. 

\begin{figure}[ht]
\begin{center}
\vspace*{-.5cm}
\parbox[b]{12cm}{\psfig{width=9cm,file=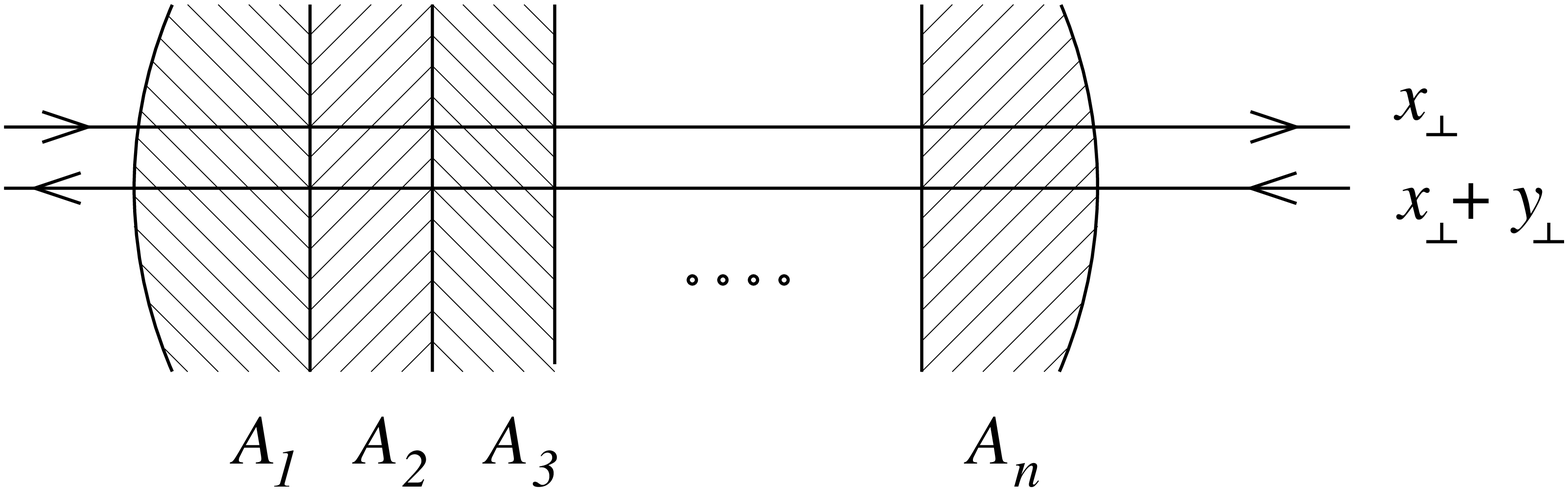}}\\
\end{center}
\refstepcounter{figure}
\label{lt}
{\bf Figure \ref{lt}:} Colour dipole travelling through a large
hadronic target.
\end{figure}

Within this framework, it is natural to introduce the additional assumption 
that the gluonic fields encountered by the partonic probe in distant regions 
of the target are not correlated. Thus, one arrives at the situation 
depicted in Fig.~\ref{lt}, where a colour dipole passes a large number of 
regions, each one of size $\sim 1/\Lambda$, with mutually uncorrelated 
colour fields $A_1$ ... $A_n$.

The crucial assumption that the fields in regions $1$ ... $n$ are 
uncorrelated is implemented by writing the integral over all field 
configurations as
\be
\int_{A}=\int_{A_1}\cdots\int_{A_n}\,\,,\label{int}
\ee
i.e., as a product of independent integrals. Here the appropriate weighting 
provided by the target wave functional is implicit in the symbol $\int_A$. 
For inclusive and diffractive parton distributions we need the
two colour contractions for products of Wilson loops,
\be\label{essence}
\mbox{tr}\left(W_{\x}(\y)W_{\x}^{\dagger}(\y)\right) 
\longleftrightarrow \frac{1}{N_c}\mbox{tr}W_{\x}(\y)\mbox{tr}
W_{\x}^{\dagger}(\y)\;.
\ee
This relation, which provides the connection between inclusive and diffractive
DIS, is the essence of the semiclassical approach.

Performing the integration over the colour fields one obtains in the large
$N_c$ limit \cite{bgh},
\bea
\int_{x_\perp}\int_A\mbox{tr}\left(W_{x_\perp}(y_\perp)
W^{\dagger}_{x_\perp}(y_\perp')\right)&=&\Omega N_c\left(1-e^{-ay^2}-
e^{-ay'^2}+e^{-a(y_\perp-y_\perp')^2}\right)\label{ww0}\,,
\\
\frac{1}{N_c}\int_{x_\perp}\int_A\mbox{tr}W_{x_\perp}(y_\perp)
\mbox{tr}W^{\dagger}_{x_\perp}(y_\perp')&=&\Omega N_c\left(1-
e^{-ay^2}\right)\,\left(1-e^{-ay'^2}\right)\,,\label{wwf}
\eea
where $\Omega = \int d^2\x$ is the geometric size of the target and $a$ plays
the role of a saturation scale. Note that according to Eqs.~(\ref{ww0}) and
(\ref{wwf}) the diffractive structure function is not suppressed by a colour 
factor relative to the inclusive structure function, in contrast to the 
suggestion made in~\cite{bh}. 

As an example, consider the inclusive quark distribution. From 
Eqs.~(\ref{dqd}), (\ref{essence}) and (\ref{ww0}) one obtains, after changing 
the integration variable $\xi$ to $N^2$,  
\be\label{idiff}
x q(x,\mu^2) = \int_x^1 d\xi {dq(\xi,b=x/\xi,\mu^2)\over d\xi}
= \int_{\x} \int_{\l} {d\ xq(x,\mu^2)\over d^2\x d^2\l}\;,
\ee
with the unintegrated quark density
\bea\label{uidiff}
{d\ xq(x,\mu^2)\over d^2\x d^2\l} &=& 
{N_c\over 32\pi^6} \int_0^{\mu^2} dN^2 N^2
\int_{\y,\y'} e^{i\l(\y-\y')}\,{\y\y'\over y\, y'}\nn\nn
&&\times\,K_1(yN)K_1(y'N)\left(1-e^{-ay^2}-
e^{-ay'^2}+e^{-a(y_\perp-y_\perp')^2}\right)\;.
\eea
This result has recently been obtained for the quark density in a large
nucleus \cite{mue} by exponentiating the amplitude for a small \qq\ pair 
scattering off a single nucleon, which is described by two-gluon exchange.
The effect of the varying thickness of the nucleus has also been taken into 
account in \cite{mue}, which makes the saturation scale $a$ dependent on the 
impact parameter $\x$.

A Glauber type model with two-gluon exchange, similar in 
spirit to \cite{got}, has recently been used to study the effect of parton 
saturation in inclusive and diffractive DIS \cite{gw}. Although perturbative 
two-gluon exchange is a higher twist effect, the contribution from the soft 
region can be used as a model for inclusive and diffractive DIS 
\cite{nz,bekw}. 

Particularly close to the semiclassical approach is the light cone hamiltonian
approach to diffractive processes \cite{hks}, which is also based on
diffractive parton densities expressed in terms of expectation values of
products of Wilson lines. For a hadronic target, modelled as a colour singlet
which only couples to one flavour of heavy quarks,  diffractive DIS is 
dominated by two-gluon exchange. In the semiclassical approach, the proton
is described by a superposition of colour fields. The role of classical
colour fields in the case of high gluon densities has first been discussed
by McLerran and Venugopalan in the case of a large nucleus \cite{mv}.

\begin{figure}
\begin{center}
\parbox[b]{7cm}{\psfig{file=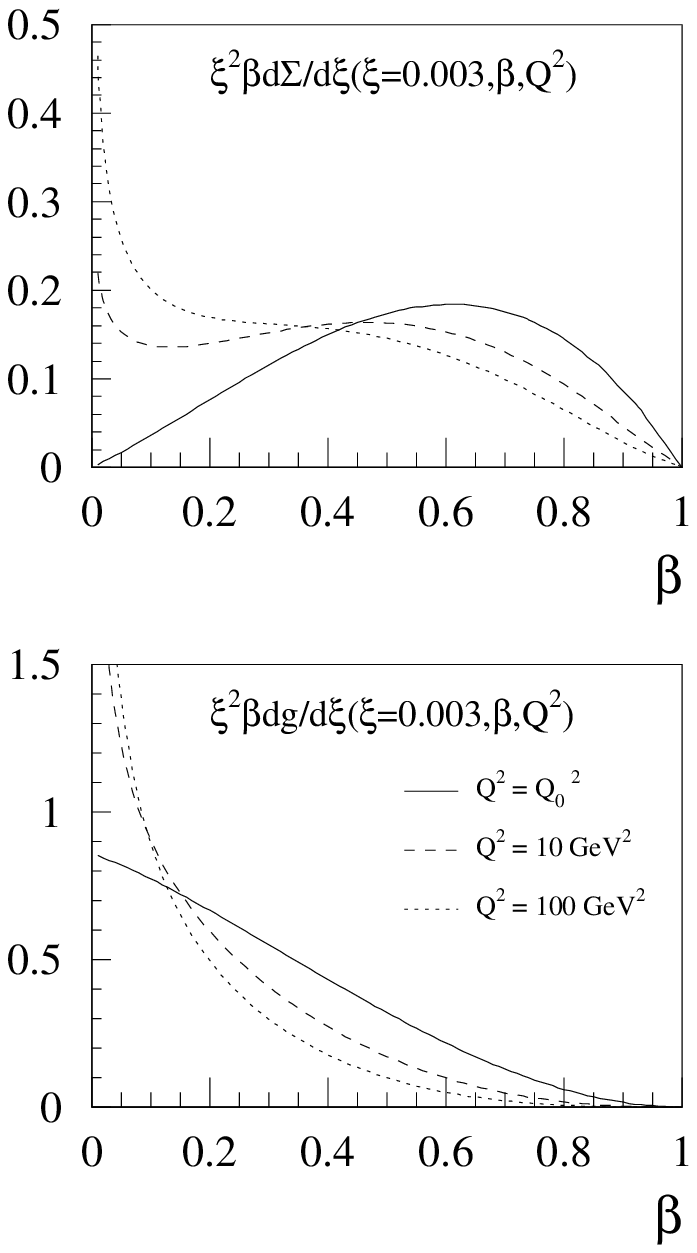,width=7cm}}\\
\end{center}
\refstepcounter{figure}
\label{figdifpdf}
{\bf Figure \ref{figdifpdf}:}
Diffractive quark and gluon distributions at the initial scale $Q_0^2$ and 
after $Q^2$ evolution. From \cite{bgh}.
\end{figure}

\subsection*{Comparison with data}\label{na}

We now use the large hadronic target as a toy model for the proton. In case 
the proton can be viewed as an ensemble of regions with independently 
fluctuating colour fields, the model might even be realistic. We have 
explicitly verified that in the semiclassical approach inclusive and 
diffractive parton distributions satisfy the DGLAP evolution equations 
\cite{dglap}. Hence, we can use the calculated quark and gluon distributions 
as non-perturbative input at some scale $Q_0^2$ and determine the 
distributions at larger $Q^2$ by means of the evolution equations. 

\begin{figure}
\begin{center}
\parbox[c]{6cm}{\psfig{file=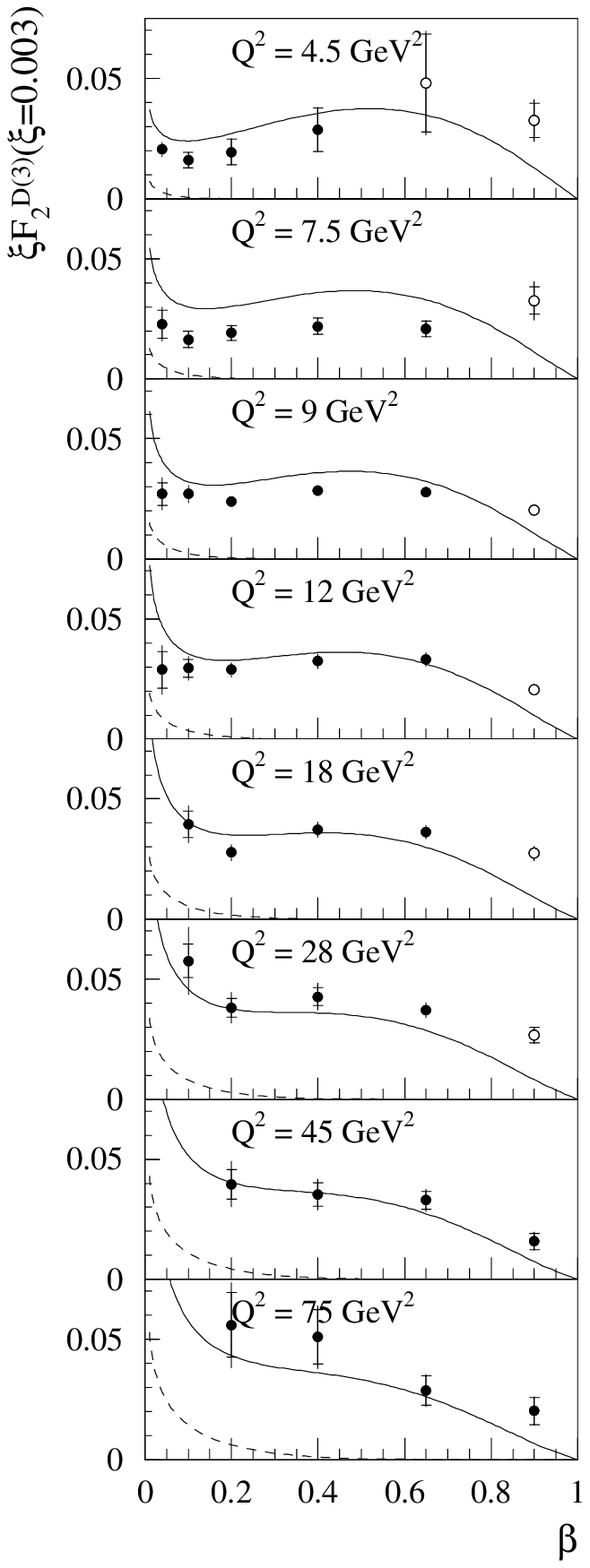,width=6cm} }\parbox[c]{2cm}{\hspace{2cm}}
\parbox[c]{6cm}{\psfig{file=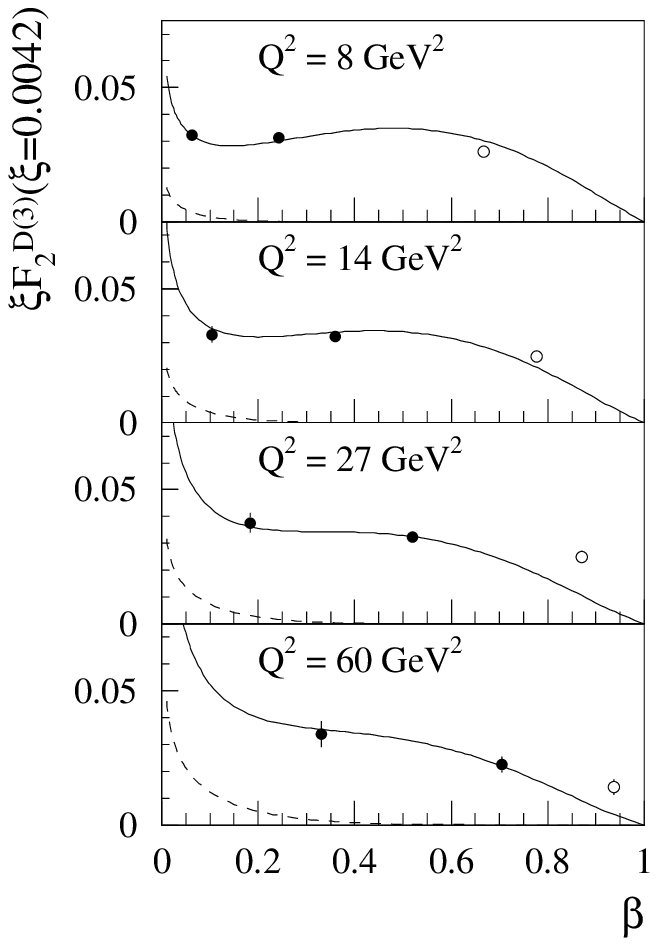,width=6cm} }\\
\vspace{0.6cm}
\end{center}
\refstepcounter{figure}
\label{figdifq2}
{\bf Figure \ref{figdifq2}:}
Dependence of the diffractive structure function $F_2^{D(3)}$ on $\beta$
and $Q^2$, compared to data from H1 (left)~\protect\cite{diffh} and ZEUS 
(right)~\protect\cite{diffz}. Open data points 
correspond to $M^2\leq 4$~GeV$^2$. The charm content of the 
structure function is indicated as a dashed line. From \cite{bgh}.
\end{figure}

\begin{figure}
\begin{center}
\parbox[b]{14cm}{\psfig{file=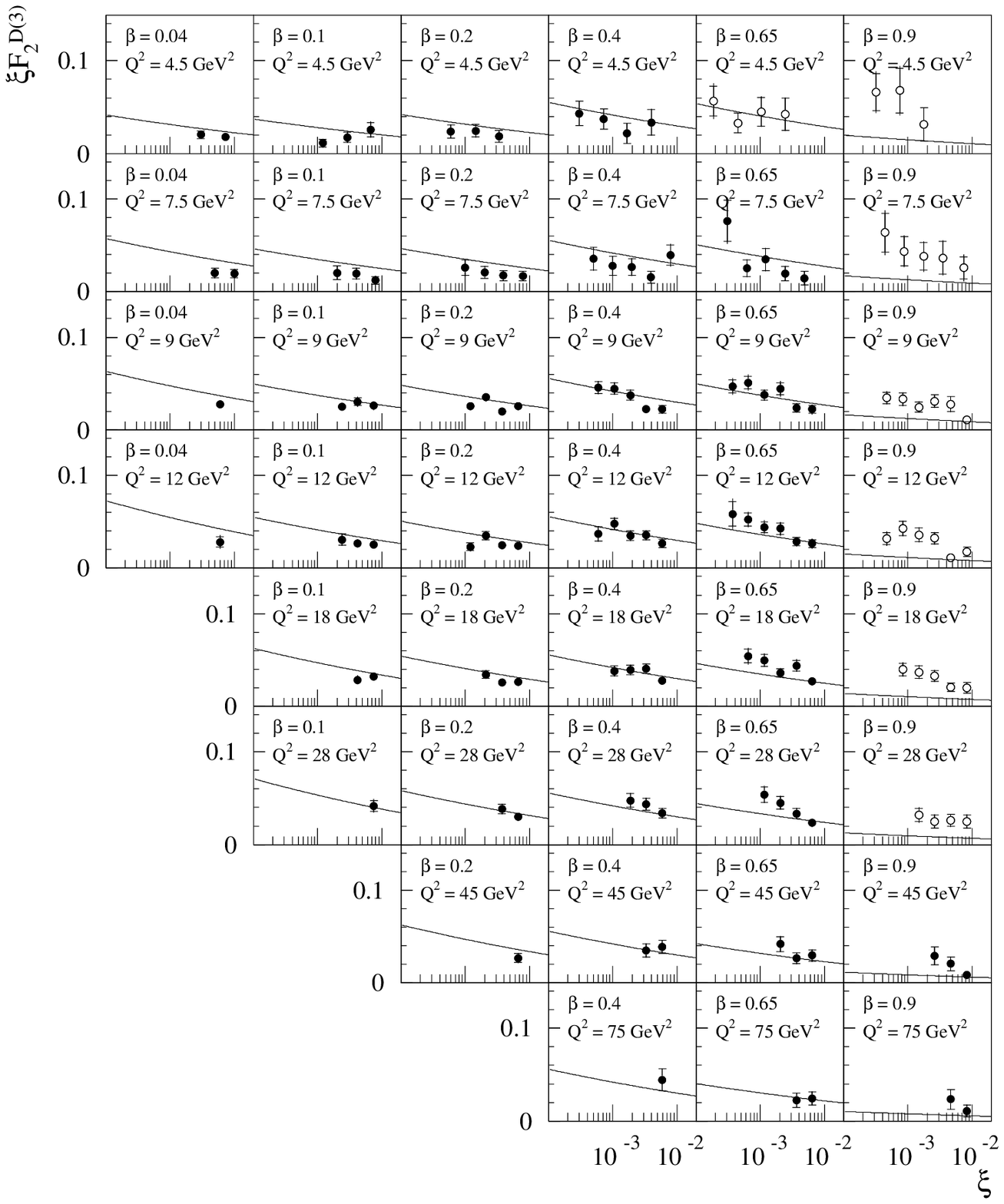,width=14cm}}\\
\end{center}
\refstepcounter{figure}
\label{figf2dh1}
{\bf Figure \ref{figf2dh1}:}
The diffractive structure function $F_2^{D(3)}(\xi,\beta,Q^2)$ at small $\xi$ 
computed in the semiclassical approach, using the fitted parameters 
given in the text.
H1 data taken from~\protect\cite{diffh}. The open data points correspond
to $M^2 \leq 4$~GeV$^2$ and are not included in the fit. From \cite{bgh}.
\end{figure}

For a given colour field, the semiclassical description of parton distribution 
functions always predicts an energy dependence corresponding to a classical 
bremsstrahlung spectrum: $q(x),g(x)\sim 1/x$. One expects that, in a more 
complete treatment, a non-trivial energy dependence is induced since the 
integration over the soft target colour fields encompasses more and more modes 
with increasing energy of the probe \cite{bgh}. At present we are unable to 
calculate this non-perturbative energy dependence from first 
principles. Instead, we choose to parametrize it in the form of a soft, 
logarithmic growth of the normalization of diffractive and inclusive parton 
distributions with the collision energy $\sim 1/x$, consistent with the 
unitarity bound. This introduces one further parameter, $L$, into the model, 
\be
\Omega \to \Omega \left(L - \ln x \right)^2.
\ee

Including this energy dependence, one obtains the following compact
expressions for inclusive and diffractive parton distributions \cite{bgh}, 
\bea
xq(x,Q_0^2) & = & \frac{a \Omega N_c \left( L - \ln x \right)^2}
{3 \pi^3} \left(\ln\frac{Q_0^2}{a} - 0.6424\right)\; , \label{qiinp}\\
xg(x,Q_0^2) & = & \frac{ 2 a \Omega N_c \left( L - \ln x \right)^2}
{\pi^2 \alpha_s(Q_0^2)}\; , \label{giinp}\\
\frac{dq\left(\beta,\xi,Q_0^2\right)}{d\xi} & = & \frac{a\Omega N_c 
(1-\beta) 
\left(L - \ln \xi\right)^2}{2\pi^3\xi^2} f_q(\beta)\; , \label{qdinp} \\
\frac{dg\left(\beta,
\xi,Q_0^2\right)}{d\xi} & = & \frac{a\Omega N_c^2 (1-\beta)^2 
\left(L - \ln \xi\right)^2}{2\pi^3 \beta \xi^2} f_g(\beta)\; \label{gdinp}.
\eea
These expressions are only applicable in the small-$x$ region, which we define
by $x\leq \xi \leq 0.01$. The functions $f_{q,g}(\beta)$ are parameter free
predictions. The model does not specify whether, in the diffractive case, the 
energy-dependent logarithm  should be a function of $x$ or of $\xi$. However, 
both prescriptions differ only by terms proportional to $\ln \beta$, which can
be disregarded in comparison with $\ln x$ or $\ln \xi$ in the small-$x$ region.

\begin{figure}[t]
\begin{center}
\parbox[b]{15.5cm}{\psfig{file=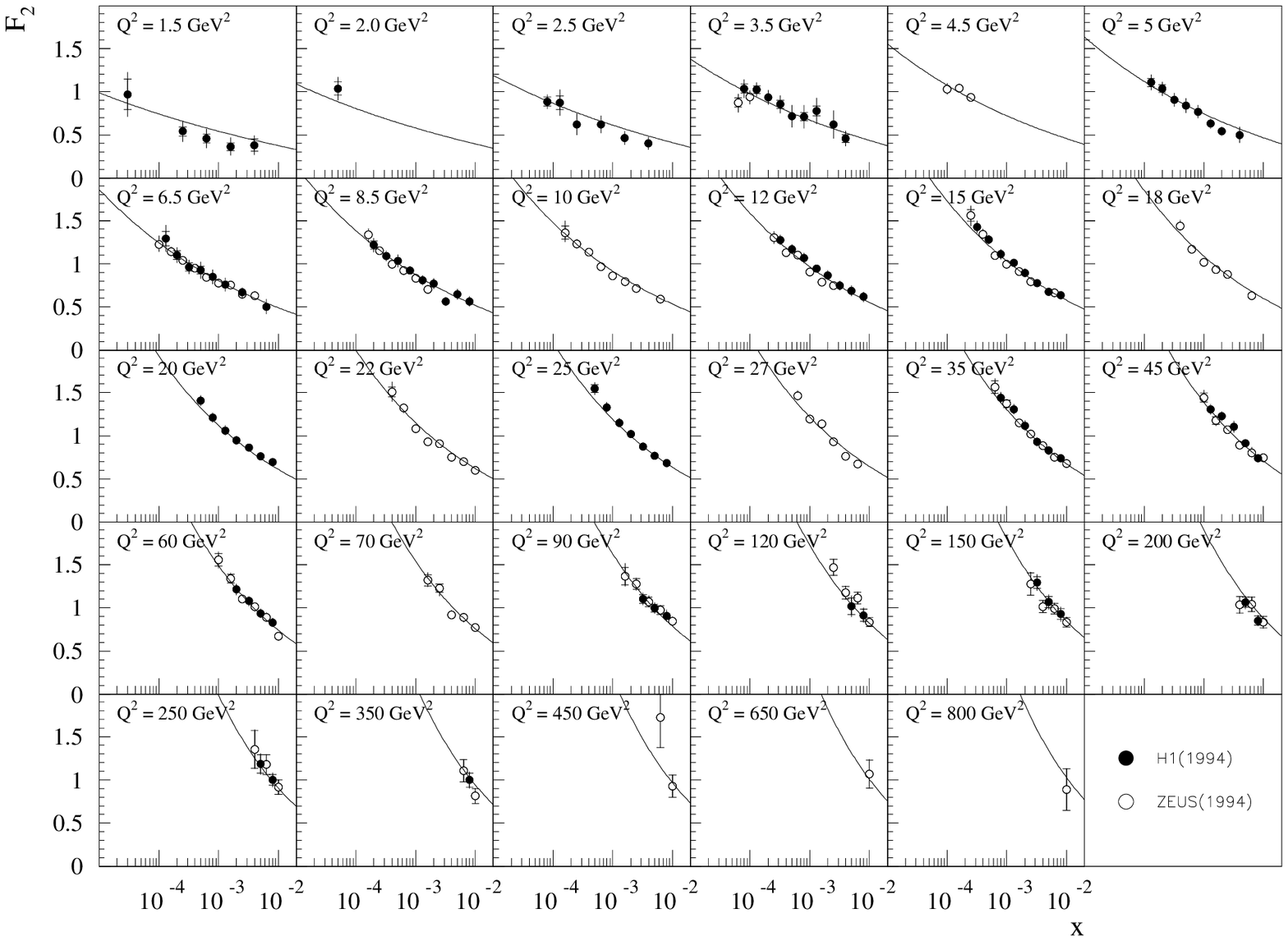,width=15.5cm}}\\
\end{center}
\refstepcounter{figure}
\label{figf2i}
{\bf Figure \ref{figf2i}:}
The inclusive structure function $F_2(x,Q^2)$ at small $x$ 
computed in the semiclassical approach, using the fitted parameters 
given in the text.
Data taken from~\protect\cite{incl}. The data with $Q^2 = 1.5$~GeV$^2$ 
are not included in the fit. From \cite{bgh}.
\end{figure}

The above equations summarize our input distributions, depending on 
$a$, $\Omega$, $L$, and the input scale $Q_0^2$. At this order, the measured 
structure function $F_2$ coincides with the transverse structure function 
$F_T$. We assume all three light quark flavours to yield the same 
contribution, such that the singlet quark distribution is simply six times the
quark distribution defined above, both in the inclusive and in the diffractive
case,
\begin{equation}
\Sigma (x,Q^2) = 6\, q(x,Q^2)\; , \qquad \frac{d\,\Sigma(\xi,\beta,Q^2)}{d\xi} 
= 6\,  \frac{dq(\xi,\beta,Q^2)}{d\xi}\; .
\end{equation}
Valence quark contributions are absent in the semiclassical approach, which 
does not account for the exchange of flavour quantum numbers between the 
proton and the fast moving virtual photon state. Charm quarks are treated 
as massive quarks in the fixed flavour number scheme \cite{ffns} (we use
$\Lambda_{{\rm LO},n_f=3}= 144$~MeV, $\alpha_s(M_Z)=0.118$, $m_c=1.5$~GeV,
$m_b=4.5$~GeV, $\mu_c = 2 m_c$). A fit to the data yields for the model 
parameters $Q_0^2 = 1.23\ {\rm GeV}^2,\ L = 8.16, 
\Omega = (712\ {\rm MeV})^{-2},\ a = ( 74.5\ {\rm MeV})^2$ . The starting 
scale $Q_0^2$ is in the region where one would expect the transition between 
perturbative and non-perturbative dynamics to take place; the two 
other dimensionful parameters $\Omega L^2$ and $a$ are both of the order 
of typical hadronic scales. 

The perturbative evolution of inclusive and diffractive structure functions 
is driven by the gluon distribution, which is considerably larger 
than the singlet quark distribution in both cases. 
The ratio of the inclusive singlet quark and gluon distributions can be 
read off from Eqs.~(\ref{qiinp}) and (\ref{giinp}). For the obtained fit 
parameters, it turns out that the inclusive gluon distribution is about twice 
as large as the singlet quark distribution.

The relative magnitude and the $\beta$ dependence of the diffractive 
distributions are completely independent of the model parameters. Moreover, 
their absolute normalization is, up to the slowly varying factor 
$1/\alpha_s(Q_0^2)$, closely tied to the normalization of the 
inclusive gluon distribution. 

Figure~\ref{figdifpdf} displays the diffractive  distributions for
fixed $\xi=0.003$ and different values of $Q^2$. The $\beta$ dependences 
of the quark and the gluon distribution at $Q_0^2$ are substantially 
different: the asymmetric quark distribution $\beta d\Sigma/d \xi$ is peaked 
around  $\beta \approx 0.65 $, thus being harder than the symmetric 
distribution $\beta (1-\beta)$ suggested in~\cite{dl}. The gluon distribution
$\beta d g/d \xi$, on the other hand, approaches a constant for 
$\beta \to 0$ and falls off like $(1-\beta)^2$ at large $\beta$. 
In spite of the $(1-\beta)^2$ behaviour, gluons remain important 
even at large $\beta$, simply due to the large total normalization of this 
distribution (the $\beta$ integral over $\beta dg/d\xi$ at $Q_0^2$ is 
approximately three times the $\beta$ integral over $\beta d\Sigma/d\xi$). 
As a result, the quark distribution does not change with increasing $Q^2$ 
for $\beta\approx 0.5$ and is only slowly decreasing for larger values 
of $\beta$. 

The dependence of the diffractive structure function on $\beta$ and $Q^2$ is 
illustrated in Fig.~\ref{figdifq2}, where the predictions are compared with 
data from the H1 and ZEUS experiments~\cite{diffh,diffz} at fixed $\xi$.
Disregarding the large-$\beta$ region, the model gives a good description of 
the $\beta$ dependence of the diffractive structure function for all values of
$Q^2$. It is remarkable that the qualitative features of the $\beta$ and
$Q^2$ dependence are also correctly described by the perturbative approach
of \cite{hks}. This indicates that the $\beta$ dependence of the diffractive
structure function is to a large extent determined by the kinematics of the
\qq\ and \qqg\ fluctuations and only partly sensitive to the details of the 
soft interaction with the proton. Also the $\xi$ dependence of the diffractive
structure function is rather well described for  $M^2 > 4$~GeV$^2$, as
demonstrated in Fig.~\ref{figf2dh1}. It has been demonstrated in \cite{bekw}
that higher twist contributions can account for the data at low $M^2$
(below 4 GeV$^2$). This is analogous to the breakdown of the leading twist 
description of the inclusive structure functions, where it occurs for similar 
invariant hadronic masses, namely $W^2\lapprox 4$~GeV$^2$ \cite{mrst2}.

Finally, also the data of the H1 and ZEUS experiments on the inclusive 
structure function $F_2(x,Q^2)$~\cite{incl} are well reproduced by the model,
as demonstrated by Fig.~\ref{figf2i}.

\section{Open questions}\label{conc}

The theoretical work, stimulated by the observation of the large rapidity
gap events at HERA, has led to a clear understanding of diffractive DIS
as a leading twist phenomenon. Diffractive parton distribution functions
can be defined completely analogous to inclusive parton distribution 
functions. Both kinds of distribution functions obey the perturbative QCD 
evolution equations. The parton distribution functions cannot be calculated
perturbatively, any relation between them reflects a non-perturbative
property of the proton. The leading twist description breaks down at
$W^2<4$ GeV$^2$ and $M^2<4$ GeV$^2$ for inclusive and diffractive DIS, 
respectively.

A physical picture for diffractive DIS, and the relation to inclusive DIS, 
is most easily obtained in the proton rest frame where all DIS processes 
correspond to the scattering of partonic fluctuations of the virtual photon 
off the proton. In the semiclassical approach the proton is described by a
superposition of colour fields. The qualitative properties of the $\beta$
dependence of the diffractive structure function are well reproduced by
the scattering of colour dipoles off colour fields, which are generated 
either by a small colour dipole or by a large nucleus. This supports our 
general ideas about diffractive DIS. It will be interesting to see whether a
more precise measurement of the $\beta$ spectrum will be quantitatively
consistent with the idea of a colour field fluctuating independently
in different sections of the proton. 

In the semiclassical approach the proton colour field is assumed to be
dominated by soft modes. Hence, diffractive and non-diffractive DIS
events are kinematically very similar at small $x$, i.e. large hadronic
energies $W$. This leads to an approximate relation between the inclusive
and diffractive structure functions \cite{bh1,b},
\be\label{scal}
F_2^{D(3)}(\xi,\beta,Q^2) \sim {1\over \ln{Q^2}} F_2(x=\xi,Q^2)\;,
\ee
which is in broad agreement with data \cite{diffz}.
For fixed momentum transfer $Q^2$ and diffractive mass $M$, both structure 
functions have the same dependence on the $\gamma^* p$ center-of-mass energy 
$W$. The factor $1/\ln{Q^2}$ reflects the suppression of small colour 
dipoles in diffractive scattering.

The dependence of the diffractive structure functions on $\xi$ is not
affected by the perturbative QCD evolution, contrary to the $x$ dependence
of the inclusive structure functions, and therefore a genuine non-perturbative
property of the proton. Hence, Eq.~(\ref{scal}) can only hold as long as the
effect of the perturbative evolution can be approximated by a single
$\ln{Q^2}$ factor. The $\xi$ dependence of the diffractive structure
function then plays the role of the non-perturbative input for the inclusive 
structure function at some low scale $Q_0^2$.

One expects that, due to unitarity, diffractive and inclusive structure
functions satisfy a relation similar to the one between elastic and
total proton-proton cross section \cite{mue1},
\bea
\sigma_{el}  &=& \int d^2b \left(1-S(b)\right)^2\;,\label{undi}\\
\sigma_{tot} &=& 2 \int d^2b \left(1-S(b)\right)\;,\label{unin}
\eea
where $S(b)$ is the S-matrix at a given impact parameter $b$. Recently, it has
been shown that this relation also holds for the diffractive and inclusive
cross sections of a \qq\ pair off the proton, if the diffractive cross section
is defined by the colour singlet projection \cite{km}. In the semiclassical 
approach, 
this relation can be read off from Eqs.~(\ref{dsdiff}) and (\ref{dsincl}) or, 
more explicitly, from Eqs.~(\ref{ww0}) and (\ref{wwf}). After integration over
$\l$ in Eqs.~(\ref{idiff}) and (\ref{uidiff}), which yields $\y=\y'$, one 
obtains for the cross sections of a \qq\ pair with size $y$,
\bea
\sigma^D_{q\bar{q}}(y) &\propto& 
\frac{1}{N_c}\int_{x_\perp}\int_A\mbox{tr}W_{x_\perp}(y_\perp)
\mbox{tr}W^{\dagger}_{x_\perp}(y_\perp) \nonumber\\
&=& \Omega N_c\left(1-e^{-ay^2}\right)^2\;,\\
\sigma^{incl}_{q\bar{q}}(y) &\propto& 
\int_{x_\perp}\int_A\mbox{tr}\left(W_{x_\perp}(y_\perp)
W^{\dagger}_{x_\perp}(y_\perp)\right) \nonumber\\
&=& 2\ \Omega N_c\left(1-e^{-ay^2}\right)\;.
\eea
Since the dependence of the saturation parameter $a$ on the varying thickness 
of the target has been neglected, the integration over the impact parameter 
$\x$ (corresponding to $b$ in(\ref{undi}),(\ref{unin})) could be carried 
out yielding the geometric size $\Omega$ as overall factor. It will be 
interesting to extend these considerations to more complicated partonic
fluctuations.

In the coming years experiments at HERA will provide detailed information
about diffractive final states, including charm and high-\pp\ jets. 
Anticipating further support of the semiclassical approach by data, 
we can hope to learn a lot about the colour structure of the proton from
a comparison of inclusive and diffractive DIS. 
\vspace{.4cm} 

The content of this progress report is largely based on recent work with
Thomas Gehrmann and Arthur Hebecker whom I thank for an enjoyable 
collaboration.

\end{document}